\documentstyle[11pt,newpasp,twoside,epsf]{article}
\def\edcomment#1{\iffalse\marginpar{\raggedright\sl#1\/}\else\relax\fi}
\reversemarginpar

\def\mic{{$\mu$m}}

\def\h2o{H$_2$O}

\def\aple{$\mathrel{\hbox{\rlap{\hbox{\lower4pt\hbox{$\sim$}}}\hbox{$<$}}}$ }
\def\apge{$\mathrel{\hbox{\rlap{\hbox{\lower4pt\hbox{$\sim$}}}\hbox{$>$}}}$ }

\begin{document}
\title{Massive Star Birth in the Inner Galaxy: Obscured Massive Star Clusters}
 \author{Robert D. Blum}
\affil{Cerro Tololo Interamerican Observatory, Casilla 603, La Serena, Chile,
rblum@noao.edu}
\author{Peter S. Conti}
\affil{JILA, University of Colorado, Campus Box 440, Boulder, CO, 80309, 
pconti@jila.colorado.edu}
\author{Augusto Damineli \& Elysandra Figuer\^edo}
\affil{ IAG-USP, Av. Miguel Stefano 4200, 04301-904, S\~{a}o Paulo, Brazil,
damineli@iagusp.usp.br}

\begin{abstract}

The current census of, and stellar population in, massive Galactic
star clusters is reviewed.  In particular, we concentrate on a recent
survey of obscured Galactic Giant H~II (GHII) regions and the associated
stellar clusters embedded in them. The regions have been selected as
the most luminous radio continuum sources, and as such the stellar
clusters appear to be among the youngest massive clusters in the
Galaxy. The emergent stellar populations are further studied through
near infrared spectroscopy of the brighter members. We also discuss
the massive stellar clusters within 50 pc of the Galactic center (GC),
comparing their known properties to those found in the GHII region
survey. It is suggested that the somewhat younger clusters associated
with the GHII regions are more suited to measuring the initial mass
function in massive star clusters. Narrow band images in the central
pc of the GC are presented which identify the young stellar sequence
associated with the evolved He~I emission line stars.

\end{abstract}

\section{Introduction}

Near infrared (1--2.5 \mic) spectroscopic classification techniques
have recently been developed for OB stars (Hanson, Conti, \& Rieke
1996; Blum et al. 1997, Hanson, Rieke, \& Luhman 1998) and Wolf--Rayet
(WR) stars (e.g., Eenens, Williams, \& Wade 1991; Figer, McLean, \&
Morris 1997). Coupled with infrared spectrometers on large telescopes,
these classification schemes are now pushing forward the exploration
of optically obscured, young stellar populations throughout the inner
Galaxy. Propelled by the pioneering work of Hanson, Howarth, \& Conti
(1997) who presented a detailed investigation of the ionizing O and B
stars in M17, we have carried out a survey of Galactic giant H~II
(GHII) regions (Blum, Conti, \& Damineli 1999, Blum, Damineli, \&
Conti 2000, Blum, Conti, \& Damineli 2001, Figuer\^edo 2001, Figuer\^edo
et al. 2002, Conti \& Blum 2002$a,b$). As summarized below, these near infrared studies are
producing a wealth of new information on the embedded stellar content
in GHII regions including the discovery of young stellar objects
(YSO), massive star formation processes, and new distance
determinations through spectroscopic parallaxes. For the purposes of
this review, we take the term YSO to include hydrogen burning objects
buried in ultra--compact H~II (UCHII) regions.

While this work has concentrated on GHII regions, great progress is
also being made on investigating the central stars of compact and
ultra--compact H~II regions (Watson \& Hanson 1997, Henning et
al. 2001, Kaper 2002) using similar techniques. As shown in the next
sections, the GHII region sample is aimed at investigating star
formation in the most massive clusters where the presence of multiple
O stars may affect both the process and resultant mass function. The
nearby Orion star forming region is then seen as a transition object
between regions of lower mass star formation and higher mass star
formation. We can expect the great body of work established in Orion
(e.g. Zinnecker et al. 1993, McCaughrean et al. 1994, Hillenbrand
1997, Hillenbrand \& Carpenter 2000) on the mass function there to
provide an important reference point to the GHII region
investigations.

The young stellar content in the Galactic center (GC) has also been
intensely studied at near infrared wavelengths. It has been revealed
that the stellar cluster in the central parsec, as well as two other
nearby clusters, are rich in OB and Wolf--Rayet stars (see Morris \&
Serabyn 1996 for a recent review). We will discuss below several
aspects of massive stars in the GC which have been the subject of
recent large telescope studies.

\section{Giant H~II Regions}

Our sample of GHII regions includes all objects in the list of Smith,
Mezger, \& Biermann (1978) for which the Lyman continuum output
indicates multiple O stars are present ($>$ 10$\times$10$^{49}$ sec$^{-1}$). The
target GHII Regions are shown in Table~1. Where available,
literature references are given for each region. 

\begin{table}
\scriptsize
\caption{Giant H~II Regions from Smith et al. (1978)} 
\begin{tabular}{llccc}
\tableline
Name&Radio Name&Distance (kpc)$^a$&N LyC (10$^{49}$)$^a$& Notes$^b$\\
\tableline
RCW49&  G284.3-0.3&     6&      96 &* \\
NGC3576&        G291.3-0.7&     3.6& 26&Figuer\^edo 2001, * \\
NGC3603&        G291.6-0.5&     8.2&    188 & Brandl et al. 1999, Eisenhauer et al. 1998\\
-&      G298.2-0.3&     11.7&   61 &*\\
-&      G298.9-0.4&     11.5&   57 \\
-&      G305.2+0.2&     8:&     49 \\
-&      G305.4+0.2&     8&      43 \\
-&      G316.8-0.1&     12.1&   80 \\
-&      G331.5-0.1&     11.1&   100& \\
-&      G333.0-0.4&     13.6:&  156&*\\
-&      G333.1-0.4&     13.4&   169&* \\
-&      G333.6-0.2&     14.1&   1140&* \\
-&      G336.8-0.0&     12.4&   192 \\
-&      G338.4+0.0&     15.3:&  208 \\
RCW122& G348.7-1.0&     $>$4.5&   14 \\
-&      G351.6-1.3&     $>$4.5&   11 &*\\
SgrA&   G0.5-0.0&       10&     132 \\
SgrB&   G0.7-0.0&       10&     308 \\
W31&    G10.2-0.3&      5.1:&   30 &Blum et al. (2001), *\\
W33&    G12.8-0.2&      4.6&    5 &*\\
M17&    G15.1-0.7&      2.3&    54& Hanson et al. (1997)\\
-&      G20.7-0.1&      18.8&   42 \\
W41&    G22.8-0.3&      12.2&   110 &*\\
W42&    G25.4-0.2&      13.4&   82 &Blum et al. (2000), *\\
W43&    G30.8-0.0&      7&      107&Blum et al. (1999), *\\
W49&    G43.2-0.0&      13.8&   172& Conti \& Blum (2002$a,b$), *\\
W51&    G49.x-0.3&      6&      154& Goldader \& Wynn-Williams (1994)\\
\tableline
\tableline

\end{tabular}

$a$ Original distance from Smith et al. (1978), using distance to the
Galactic center (GC) of 10 kpc. Distance and luminosity should be
revised for a distance to the GC of 8 kpc ( Reid 1993). New near
infrared spectroscopic distances have been determined for M17, W43,
W42, and W31; see text. A ``:'' indicates the far radio recombination
line distance.

$b$ Reference for near infrared observations. A ``*'' indicates $JHK$
imaging and/or spectroscopy exist from the present survey.

\end{table}

\begin{figure}
\plotone{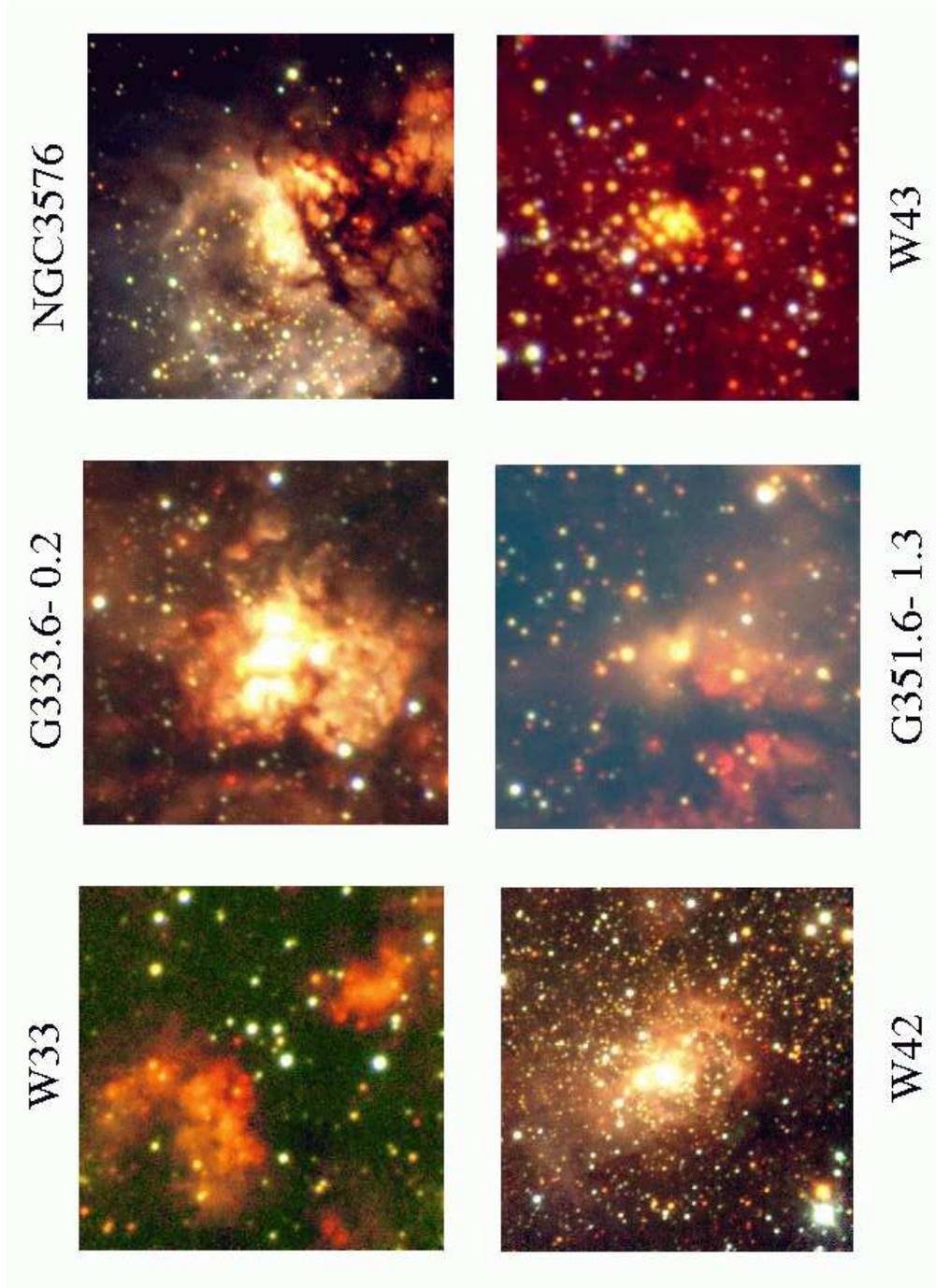}
\caption[] {A representative sample of $JHK$ three--color images for
survey clusters. The clusters can serve as a loose ``evolutionary''
sequence. The OB stars in W33 and G333.6-0.2 are still highly veiled,
while those in W43 have emerged and show normal infrared spectral
types; see text.}
\end{figure}

Near infrared imaging has revealed dense, rich clusters of new born
stars in nearly all of the GHII regions listed in Table~1.  Figure~1
shows a representative subset of the clusters imaged to date. The
majority show a combination of complex nebular emission, regions of
high and variable extinction, and centrally concentrated clusters of
stars. A notable exception is W49.  Only ultra--compact H~II regions
are seen toward the core of W49, with several near infrared
counterparts. W49 is discussed at length by Conti \& Blum 2002$a,b$.
Figure~1 may be thought of as a rough ``evolutionary'' sequence in the
sense that the GHII regions to the upper left (W33, G333.6-0.2) are
still very much embedded and so probably younger. The near infrared
point sources are still highly veiled by the hot dust from their birth
cocoons that no photospheric features have been detected in the
candidate O stars. To the lower right (e.g. W43) the clusters have
become more revealed and individual stars have well determined
spectral types. The brightest object in W43 is a Wolf--Rayet type star
(WN7) which suggests an age \apge 2 Myr (Blum et
al. 1999). Intermediate clusters such as W42 and W31 have main
sequence O stars (Blum et al. 2000, 2001) which are very young. It is
clear that selecting GHII regions by their Lyman continuum output
generally biases the survey to the youngest emergent clusters since
the associated nebulosity has not yet been dispersed by the energetic
winds and radiation pressure from the hot stars. This means that
somewhat older, even luminous clusters could go undetected.

\begin{figure}
\plottwo{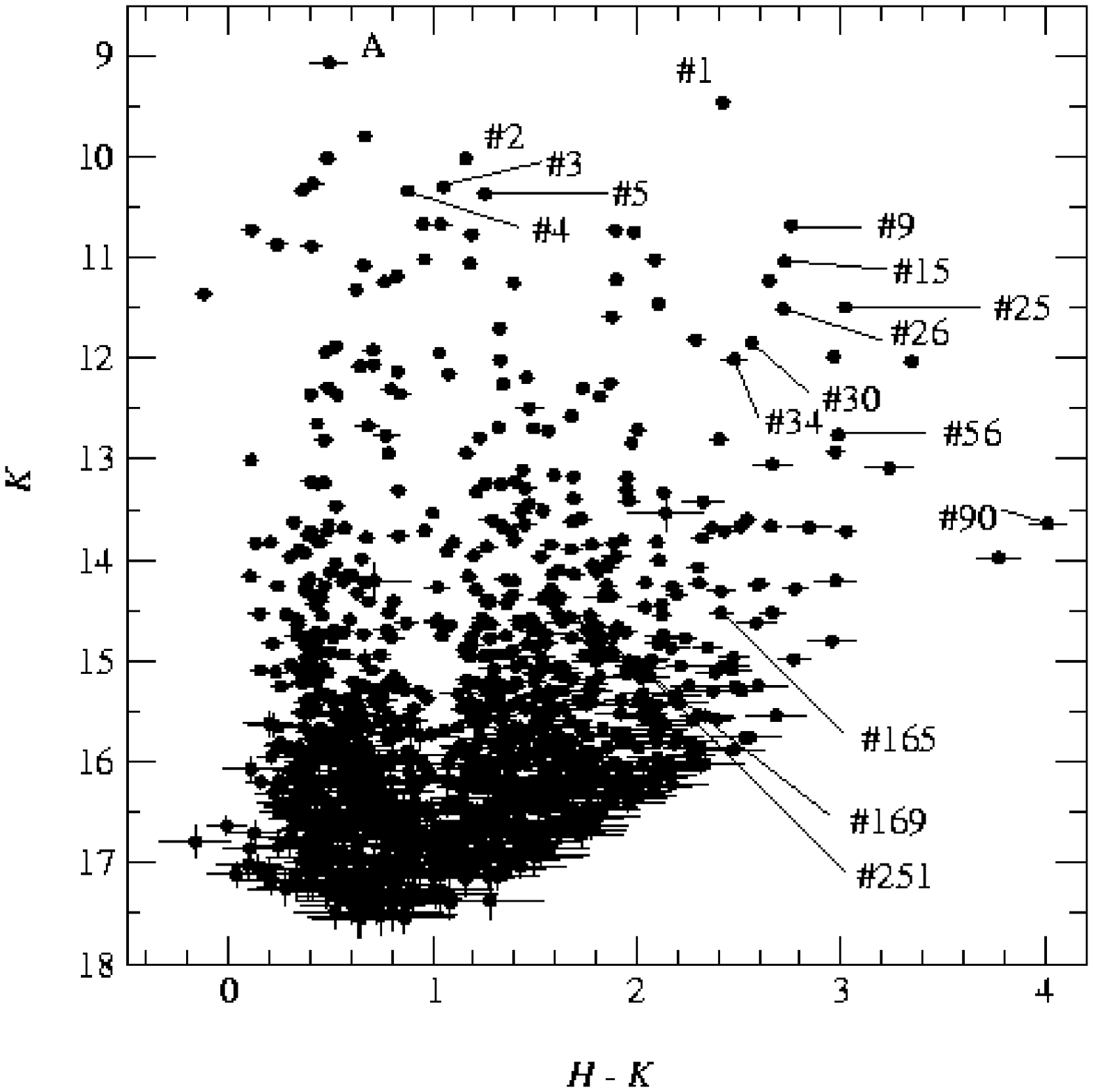}{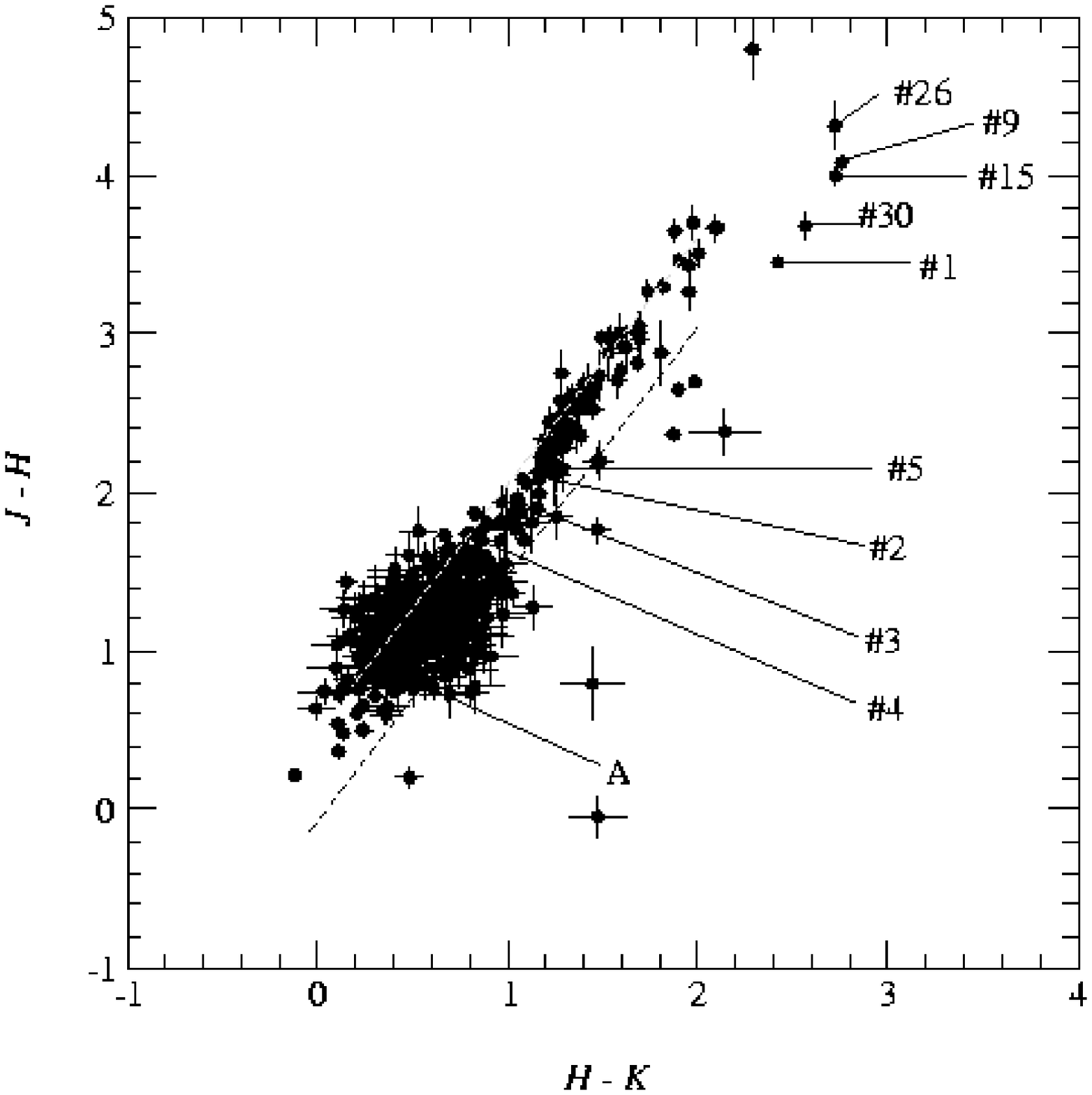} 
\caption[]{$H-K$ color--magnitude diagram and $J-H$ vs. $H-K$
color--color plot for the W31 cluster and surrounding field. The stars
labeled \#2 -- \#5 are O stars. Stars labeled \#1, 9, 15, 26, and 30
are massive young stellar object candidates;.  The remaining labeled
objects are candidate counterparts to 5 GHz radio sources.}
\end{figure}

A typical color magnitude diagram (CMD) and color--color plot are
shown in Figure~2. These particular diagrams are for W31, but the
basic features are common to the GHII region clusters in general. The
main features are a foreground sequence at bluer $H-K$, a cluster
sequence to the red, strong differential reddening which produces a
larger scatter in $H-K$ than the typical photometric uncertainty, and
a sequence of stars with an indicated excess of emission in $H-K$ in
the color--color plot. These objects lie to the red in this diagram
compared to stars whose colors are consistent with ``normal'' stellar
colors seen through a column of dust (some combination of interstellar
and local). 

The presence of these ``excess'' objects is particularly exciting
because it allows us to investigate aspects of the massive star birth
process; their presence also strongly suggests that revealed O stars
are on, or nearly on the zero age main--sequence (ZAMS, Hanson et
al. 1997, Blum et al. 2000, 2001). For all the clusters associated
with GHII regions in Table~1 for which a $J-H$ vs. $H-K$ diagram
exists and for which some massive stars are conclusively identified by
spectroscopic means, there exist young (e.g. UCHII) or pre main
sequence objects, or both. It appears that the hottest O stars in M17
(Hanson et al. 1997) have blown away their natal material, while the
less massive later O and B stars show clear disk signatures.  These
two groups are spatially segregated in M~17. Are the early O stars
more efficient at removing their circumstellar material, or do the two
groups represent sequential star formation? W31 has a YSO which is
brighter than the early O stars. Its spectrum exhibits permitted Fe~II
emission which Blum et al. (2001) take as evidence of a dense
circumstellar flow or disk. Accounting for the larger circumstellar
extinction and excess emission for this star, Blum et al. show
(assuming the excess arises in a disk geometry) that it is most likely
consistent with a late type O or early B star. This object is also
associated with an UCHII radio source (Ghosh et al. 1989). The Lyman
continuum output derived from the radio emission is consistent with
the late O early B classification. Thus there is very strong evidence
that some massive stars do form by processes which include a disk
accretion phase.

However, no mid to early O star (i.e. one of the most massive type)
has been found in any of the youngest clusters which shows evidence of
a circumstellar disk. All such stars have formed recently in the
presence of somewhat lower mass OB stars, some of which show
unmistakable signs of disks. It is possible that the earliest stars
are simply more efficient in removing their circumstellar material,
and the disk phase is thus shorter. On the other hand, if the
timescale for formation of the massive stars is similar to that for
the lower mass stars as has been recently suggested (within 10$\%$,
Behrend \& Maeder 2001) and these stars have formed at the same time,
then the observations might suggest that the most massive stars do not
form with associated disks.

The hot star spectra obtained in these young clusters are not just
useful for studying the star birth process. With suitable
calibrations, they can be used to determine spectrophotometric
distances to the GHII regions effectively probing Galactic
structure. The details of our technique are given in Blum et
al. (2001). Briefly, the infrared spectra are used to determine an
associated spectral type and absolute magnitude. The apparent
brightness and extinction are known from the $JHK$ photometry and a
distance is determined from the intrinsic and observed brightness. The
largest uncertainty is due to the intrinsic scatter in the known
brightnesses of the O stars (Vacca, Shull, \& Garmany 1996) and the
unknown age of the O stars. The former can be improved upon by
maximizing the number of stars in a cluster with individual distances
determined, the latter by observing fainter B stars which can't have
evolved off the main sequence appreciably.

\section{The Galactic Center}

Forrest et al. (1987) and Allen et al. (1990) discovered that a very
young stellar component ($<$ 10 Myr) exists in the Galactic center
(GC) by associating near infrared emission lines which arise in the
energetic winds of massive stars with compact (stellar) sources in the
central pc. Later, Krabbe et al. (1991) established that a cluster of
such evolved stars was located in the GC which represented the most
recent episode of star formation there. Since then, a host of
additional studies have refined the observed properties of the
emission--line stars (Najarro et al. 1994, Libonate et al. 1995, Blum
et al. 1995$a,b$, Krabbe et al. 1995, Tamblyn et al. 1996, Najarro et
al. 1997, Paumard et al. 2001). However, questions still remain about
whether the emission--line stars represent normal stellar evolution or
the result of environmental effects in the extreme GC environment
(e.g. collisional mergers). See Tamblyn et al. (1996) and Najarro et
al. (1997) for differing points of view.

\begin{figure}
\plotone{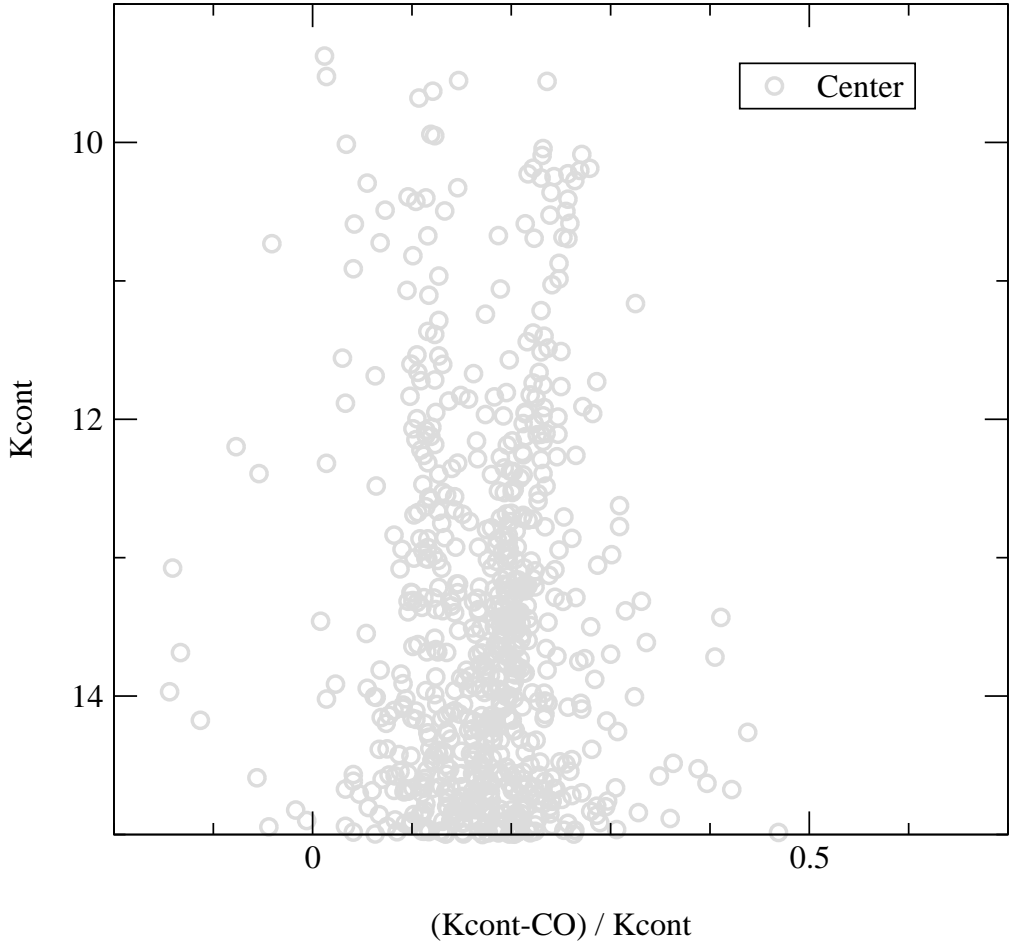}
\caption[] {Kcont magnitude vs. the CO index (Kcont flux $-$ CO flux /
Kcont flux) for the central 20$''$ of the Galactic center ({\it open
circles}).  The sequence to the left, with weak CO indices, represents
the OB stars associated with the most recent burst of star formation
in the GC.  The sequence to the right is M supergiants and older AGB
stars, which have strong CO indices that increase for brighter Kcont
magnitudes.
\label{gc1co}
}
\end{figure}

For many years, GC investigators have sought evidence of the
lower--mass stars (main--sequence and giants) which must have
accompanied the emission--line stars at the time of formation if the
latter indeed sprang from normal stellar progenitors. However, the
combined effects of crowding and differential extinction have rendered
searches for this component fruitless, until recently. Progress has
been made on this issue by combining the Gemini adaptive optics image
quality with a set of very closely spaced narrow filters. Using data
from filters centered on the CO band head at 2.3 \mic \ and a nearby
continuum position (2.26 \mic), taken as part of the Gemini demo
science program, we have identified the lower--mass sequence of stars
associated with the cluster of massive emission-line
stars. Figure~3 shows the CO index (Kcont flux $-$ CO flux /
Kcont flux) for the central 20$''$, which includes the emission-line
star cluster (IRS 16, Krabbe et al. 1991). Two sequences are
present. The older population has steadily increasing CO index (more
positive index in the Figure for stars with brighter Kcont magnitudes;
e.g, Kleinmann \& Hall 1986, Blum et al. 1996, Ram\'{\i}rez et al. 1998)
which is expected for AGB stars and M supergiants, while the young
sequence has CO indices which reflect the (essentially constant)
continuum slope of the reddened (A$_K$ $\sim$ 3) Rayleigh--Jeans tail
for hot stars.  Figure~4 shows the same index, but for stars
in a field 20$''$ N of the central field, clearly indicating the young
sequence is highly concentrated in the central region. The brightest members
of the young sequence include the IRS 16 stars.

\begin{figure}
\plotone{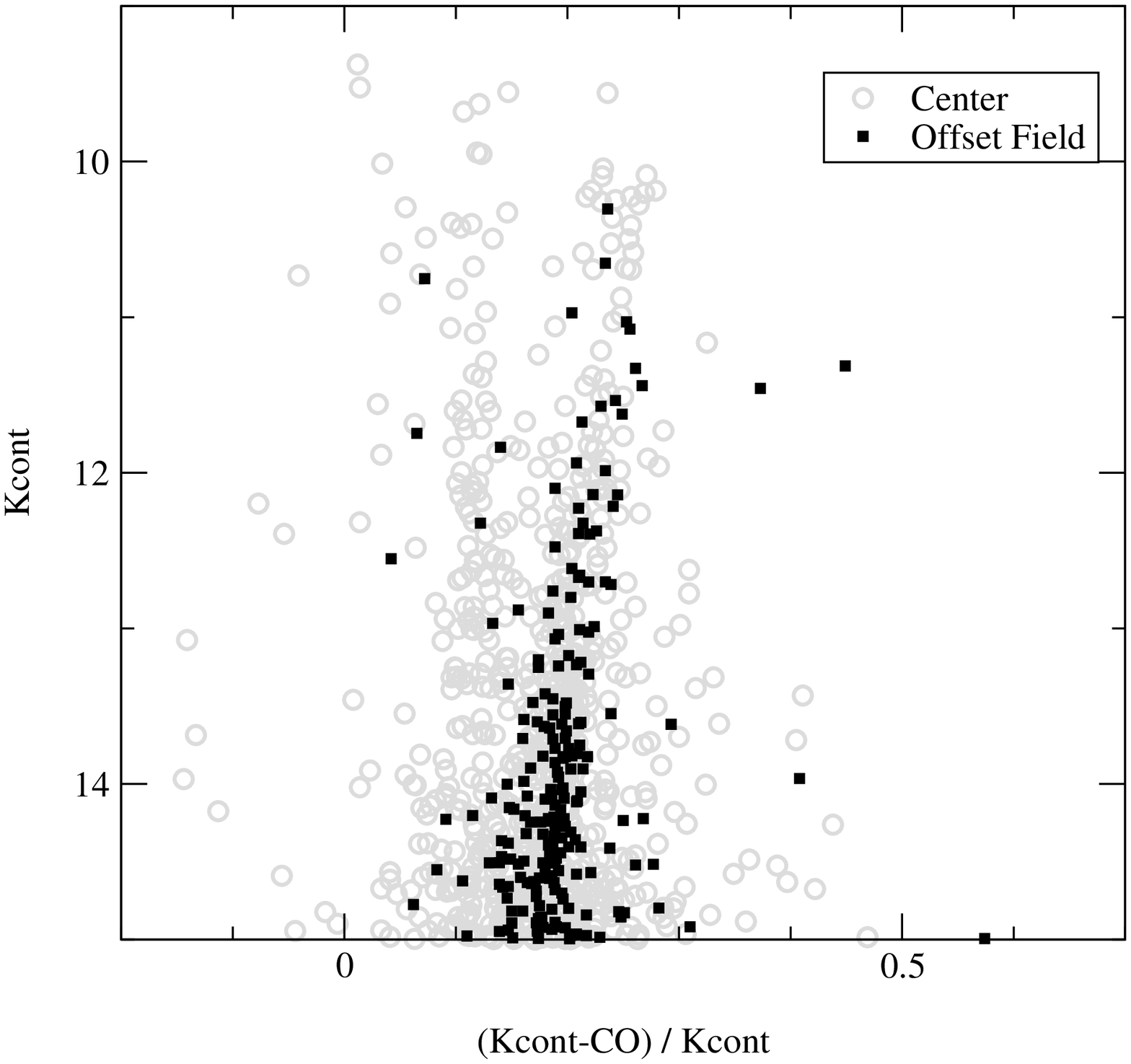}
\caption[] { Overlay of the GC CO indices with the CO indices of a
control field 20$''$ N of the GC ({\it filled squares}).  The control
field stars fall along the same CO index sequence as the older
population in the GC with strong CO indices, but there are few stars
on the young sequence which is highly concentrated to the GC field.  }
\label{gc2co}
\end{figure}

By using observations in an offset field such as shown in Figure~4, a
statistical estimate of the total number of stars associated with the
young sequence can be made in the central pc. However, spectroscopy is
still required to pin down the true nature of the young
stars. Detection of normal photospheric features in the $K-$band would
unambiguously show the young emission--line stars in the GC to be
derived from more or less normal stellar progenitors. An independent
estimate of the overall mass scale for the young stars would also
follow. Models have been computed for several of the emission--line
stars (Najarro et al. 1997) allowing them to be placed in the HR
diagram, but the results may depend on the specific physics included
(e.g, so far no line--blanketed models have been computed).  In any
case, estimating masses from the spectral types of stars based on
photospheric features is more straightforward than interpreting the
spectra of evolved emission--line objects. The young stellar sequence
indicated in Figure~4 is not confined to the so--called SgrA*(IR) cluster
immediately surrounding the black hole at the GC (Ghez et al. 1998,
Genzel et al. 2000), but includes at least some of those stars. The SgrA*(IR)
cluster stars observed spectroscopically do not show strong CO
absorption (Eckart et al. 1999, Figer et al. 2000).

\clearpage

\section{The Arches Cluster and the IMF}

The Arches star cluster was discovered recently in limited near
infrared surveys (Nagata et al. 1995; Cotera et al. 1996) 30 pc in
projection from the Galactic center (GC). Soon after, it was
recognized that the Arches is one of the most massive young star
clusters in the Milky Way and Magellanic Clouds (Serabyn et
al. 1998). Serabyn et al. estimated that the Arches contains some 100+
OB stars in a projected area of about 0.5 pc. The Arches had lain
hidden for so long only because it was obscured by 30 magnitudes of
intervening visual extinction. The Arches is one of the three nearest
young, massive star clusters which can be studied at high angular
resolution (along with NGC3603 in the Galaxy, and R136 in the Large
Magellanic Cloud-LMC). It is the only nearby cluster which is also
found in a dense circumnuclear environment. Detailed studies of such
nearby mini-starbursts are essential to establish the stellar mass
function which is produced by this prolific mode of star formation.

Following the work of Serabyn et al., Figer et al. (1999) used
HST/NICMOS images to show that the Arches is similar in total mass to
the mini-starburst cluster R136 in the LMC (\apge 10$^{4}$ M$_{\sun}$)
and perhaps 10 times as dense (3$\times10^{5}$ M$_{\sun}$
pc$^{-3}$). More importantly, Figer et al. (1999) estimated the mass
function in the Arches to be significantly flatter than a normal
Salpeter (1955) power-law.  Figer et al. find the power--law slope for
the Arches to be $\Gamma = -$0.7 compared to $-1.35$ for
Salpeter. This is in stark contrast to R136 (Massey \& Hunter 1998)
and the OB associations in the Milky Way (Massey et al. 1995) which
exhibit Salpeter--like slopes, $\Gamma =$ $-1.0$ to $-1.4$.

Figer et al. (1999) argued that the flat mass function in the Arches
was to be expected because the preconditions (namely high turbulent
velocity and magnetic field strength) in GC star--forming clouds would
naturally tend to produce more massive stars $-$ a higher Jeans mass
would be required to overcome these forces which oppose gravitational
collapse (see Morris and Serabyn 1996). But Portegies~Zwart et
al. (2001) have computed dynamical evolution models of the Arches
against the background potential of the GC. They claim the Arches mass
function may only appear flat because dynamical segregation has
removed lower mass stars to larger radii than were observed by Figer
et al. (1999).  The models can produce a flat mass function in the
core of the Arches cluster, similar to the observed mass function, but
a mass function for the entire cluster that is normal.  It is only the
observational constraint of having excluded the outer regions of the
cluster which leads to the observed mass function being anomalously
flat.

\begin{figure}
\plotone{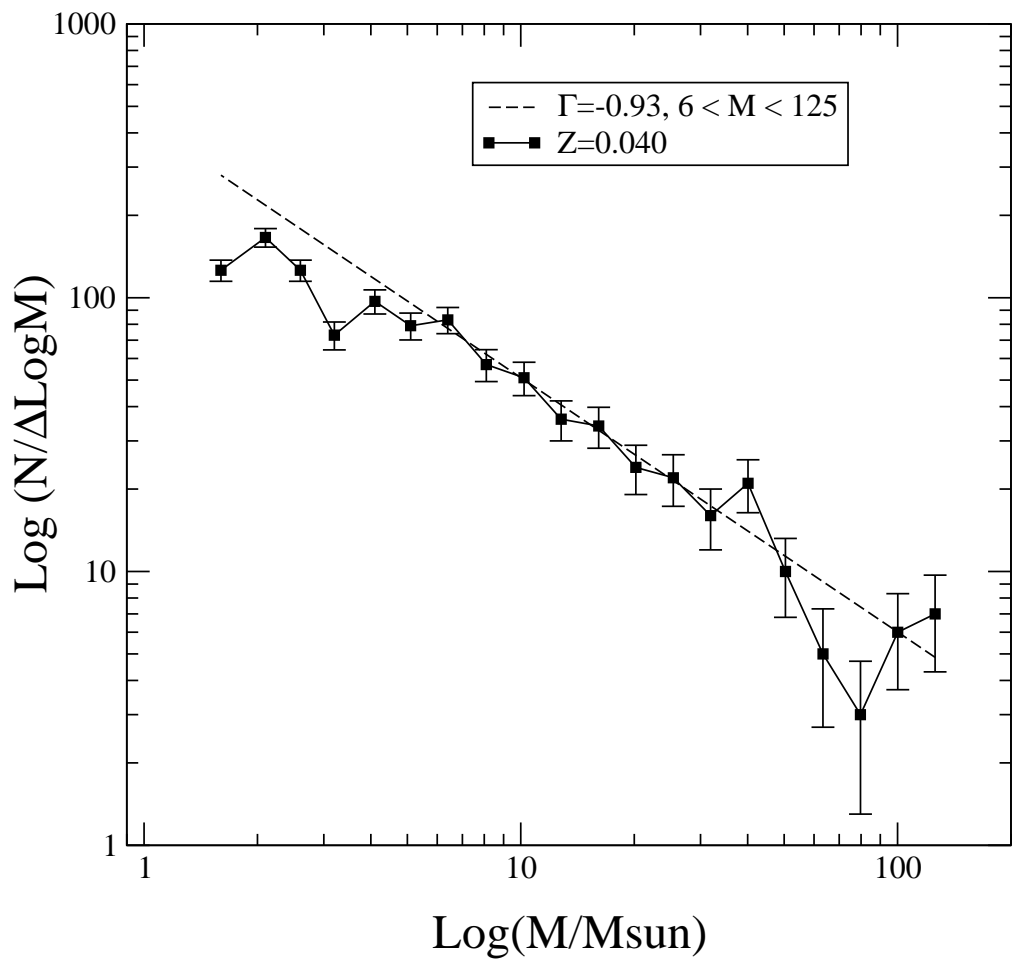}
\caption[] {Mass function in the Arches cluster derived from Gemini
North Demo Science Hokupa'a+QUIRC $K'$ images. The slope is not as
flat as the HST images, but still flatter than Salpeter (1955). The
slope is uncertain mainly due to the choice of isochrones ($\pm$
5-10$\%$) used to transform $K'$ to mass.}

\end{figure}

We have used the Gemini Demo science observations to explore the mass
function in the core of the Arches. These Gemini
adaptive optics observations have better image quality and thus go
deeper in the core of the Arches than previous HST data. Like the HST
data, the Gemini data indicate a flatter mass function than the
Salpeter (1955) case and that in R136 (Massey \& Hunter 1998). However
it is intermediate, $\Gamma$ $\approx$ $-0.9$, and not too different
from the OB associations in the Milky Way (Massey et al. 1995), but
see Eisenhauer et al. (1998) for the case of NGC3603 (who give an
upper limit to the slope, $\Gamma$ $\leq$ $-$0.73). Figure~5 shows the
mass function derived from these $K'$ images. The slope can change by
$\sim$ $\pm$ 5--10 $\%$ for different isochrones used to transfer the
$K$ magnitudes to mass (Schaller et al. 1992, Meynet et al. 1994). The
result is preliminary; there are still significant areas of
uncertainty. For example, the background stellar population (here
estimated from the HST/NICMOS data of Figer et al. 1999) is rather
uncertain. Binary fraction has not been included, and the stellar
models do not include the effects of rotation which are probably
significant (Maeder \& Meynet 2000, Meynet \& Maeder 2000). Cotera et
al. (2001) will explore these issues in more depth. Furthermore,
additional observations are needed to explore the issue of dynamical
segregation.

In the meantime, a comparison to preliminary measurements of the mass
function in the GHII region clusters places the Arches firmly at the
flat end of the distribution. The values of $\Gamma$ for W31, NGC3576,
and W42 are $-1.3$, $-1.4$, and $-1.5$, respectively. These are the
youngest clusters in our sample with identified young stellar objects
or UCHII regions. The young age and fact that these clusters are far
from the powerful influence of the inner Galactic potential suggests
they are ideal for measuring the initial mass function in massive
star, star clusters.
 
\section{Summary}

The observational basis for the emergent star birth properties of
massive stars in clusters is growing rapidly through the application
of near infrared techniques to individual clusters in the
Galaxy. Clusters with a few to more than 100 OB stars have been
identified and detailed spectra obtained for the brighter
members. Progress is being made on measuring the mass function
produced by these ``mini-starburst'' episodes. The clusters located in
the more extreme region of the Galactic center maybe the most massive
and dense clusters in the Galaxy. However, the clusters seen toward the
GHII regions may be the most ideal for establishing the mass function
since they are younger and less affected by the Galactic center
potential.

Young stellar objects (YSO) are found in essentially all these young
massive star forming regions. The analogous high mass objects to the
lower mass YSOs are ultra-compact H~II regions: OB stars which are
burning hydrogen, unlike low mass YSOs, but which have not yet
revealed their photospheres. A number of these, as well as more
revealed OB stars, show evidence for circumstellar disks. Thus, at least
some massive stars appear to form through a process which includes a
disk accretion phase. However, the most massive O stars revealed in
the young clusters do not show evidence for disks. The disk phase may
be too short to observe in the most massive stars, or perhaps it does
not occur.

For the first time, a clear 
sequence of lower mass stars has been associated with the
Galactic center He~I emission--line stars. This sequence is evident in high 
angular resolution adaptive optics narrow--band images of the central pc. 
The sequence is spatially extended ($\sim$ 20$''$) compared to the SgrA*(IR) 
cluster of stars immediately surrounding the nuclear black hole.

\end{document}